# Passively Adaptive Radiative Switch for Thermoregulation in Buildings

Charles Xiao, Bolin Liao, and Elliot W. Hawkes

**With the ever-growing need to reduce energy consumption, building materials that passively heat or cool are gaining importance. However, many buildings require both heating and cooling, even within the same day. To date, few technologies can automatically switch between passive heating and cooling, and those that can require a large temperature range to cycle states (>15º C), making them ineffective for daily switching. We present a passively adaptive radiative switch that leverages the expansion in phase-change energy storage materials to actuate the motion of louvers and can cycle states in less than 3º C. The black selective-absorber louvers induce high heat gain when closed, yet when open, expose a white, infrared emissive surface for low heat gain. During an outdoor test in which temperature was held steady, our device reduced the energetic cost of cooling by 3.1x and heating by 2.6x compared to non-switching devices. Our concept opens the door for passively adaptive thermoregulating building materials.**

## INTRODUCTION

Thermal regulation of buildings is a pressing issue and will only become more important in the coming years. Climate change is not only increasing the importance of cooling, but it will also increase the demand for heating in some areas, due to more extreme temperature fluctuations[1]. For example, the United States is expected to increase its cooling needs in every region, even as its heating needs remain high or even increase in some regions (Fig. 1A). Moreover, improved cooling and heating technologies present one of the greatest opportunities for decarbonization. Currently, 50% of the energy consumed by buildings in the US comes from cooling and heating[2]. And to achieve net zero by 2030 requires a further ~300 MT reduction in carbon dioxide from announced pledge targets[1], even as population growth and industrialization increase energy demand[3]. Developing efficient cooling and heating technologies will not only help us meet this growing demand, but also help lead us to a net zero future.

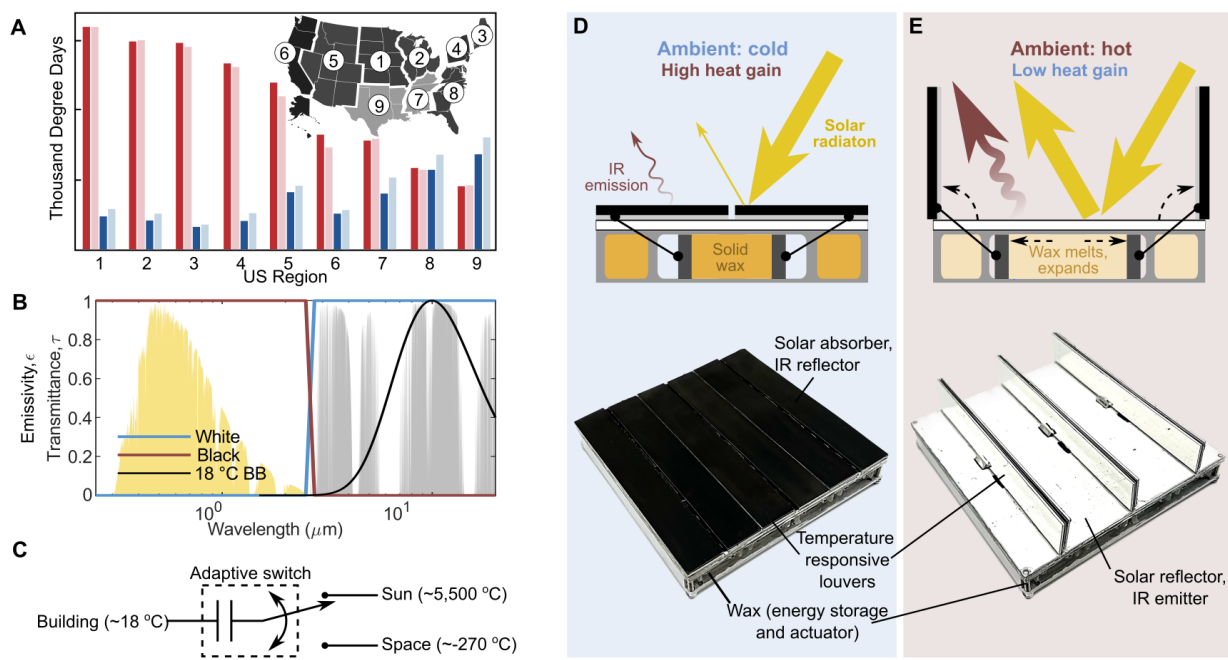

**Figure 1: The increasing heating and cooling demands could be partially addressed by the concept of a passively adaptive radiative switch for thermoregulation.**

(A) Heating degree days (red) and cooling degree days (blue) for 2020 (dark colors) and 2050 (light colors) in different United States regions[1]. Heating degree days and cooling degree days are the number of degrees that the average temperature is below or above, respectively, 65 °F (18.3 °C), summed across all days of a year. In all regions, cooling demands will increase while heating demands remain high and even increase in some areas.

(B) The desired spectral properties of our thermoregulating switch: a black, heating state (red line) should have high absorbance in the solar spectrum yet low radiance in the infrared spectrum; a white, cooling state (blue line) should have low absorbance in the solar spectrum yet high radiance in the infrared spectrum. This is enabled because of the substantial separation of the emissive spectrum of the ~5500 °C sun (yellow) and a ~20 °C object on earth (black line). Also shown is the atmospheric transmittance (gray).

(C) Our thermoregulating switch concept can be modeled as a single-pole double-throw radiative switch with a thermal capacitor that is created by the phase-change material (PCM) used. The roof's temperature is compared to the target setpoint temperature to determine when a switch should occur, independent of the ambient temperature.

(D) The physical implementation of the concept comprises louvers with a black, low IR-emissive coating on their top surface and a reflector on their bottom surface along with a base that is white and IR-emissive. The louvers are driven by the expansion of a PCM. When the device's temperature is below the setpoint, the PCM is low-volume and the louvers are closed, resulting in high heat gain. Substantial solar radiation is absorbed, and little infrared radiation is emitted.

(E) When the device's temperature is above the setpoint, the PCM expands and drives the louvers open, resulting in low heat gain. Solar radiation is reflected, and large amounts of infrared radiation is emitted to deep space. Images of the device are shown at bottom.

One promising direction for both cooling and heating is to use thermal radiation to harness the large thermal reservoirs in space: deep space (~-270 °C) and the Sun (~5,500 °C). Two factors make this possible. First, there is a spectral separation between solar radiation (250-2500 nm), which transfers heat from the sun to objects on Earth (heating the object), and mid-infrared radiation (2-50 μm), which transfers heat from objects on Earth to deep space (cooling the object) (Fig. 1B). This separation makes it possible to engineer materials that reflect in one spectrum and emit/absorb at the other. Second, there are atmospheric windows that allow certain bands to pass. The window between 7-14 μm is important for cooling because it allows for a significant portion of emitted infrared light on Earth to travel directly into space. Materials with high solar reflectivity (absorbing little heat from the sun) and high infrared emittance (transferring substantial heat to deep space), known as radiative cooling materials, exploit this window to cool below ambient temperatures. Recently, the development of such materials has attracted significant interest[4–11]. Conversely, materials with low solar reflectivity (absorbing much heat from the sun) and low infrared emittance (transferring little heat to deep space), known as selective absorbers, show efficient solar heating. This, too, has attracted a great deal of attention[12,13].

However, in many real-world buildings, both heating and cooling may be needed at different times of year or even within the same day. To achieve efficient thermal management, it is ideal for buildings to switch which of the extraterrestrial thermal reservoirs (sun or deep space) it is transferring heat with as the temperature of the building changes. Conceptually, this requires the single input of the building temperature (whether it is above or below the ideal setpoint temperature) to switch both the solar and infrared emissivities of the building material. This behavior can be modeled as a single-pole double-throw (SPDT) radiative switch, analogous to SPDT switches in electronics (Fig. 1C). Several groups have attempted such designs, but many of these designs exhibit poor spectral switch ratios[14] or are only single-throw in nature[15–21]. In our search, we found few high-performance active SPDT switches[22–24], however, these all require active switching with externally supplied power. Passive switching has the potential for enabling a low-complexity, low-cost, reliable system. Recently, two groups have discussed this concept without experimental implementation[15,25], and one research group has built a functioning device[26,27]. The device is based on the rolling of a two-way shape memory polymer. However, this design has a high switching band (the number of degrees to switch state from 10% to 90%; see Fig. 4) and large hysteresis (number of degrees between heating and cooling state curves), both greater than 15 °C. This makes it impractical for use in regulating daily temperature fluctuations.

## CONCEPTUAL OVERVIEW

Here we present a passive SPDT radiative switch for thermal regulation with: i) a low switching band, ii) small hysteresis, iii) integrated thermal energy storage, and iv) a configuration that allows nearly any heating and cooling materials to be incorporated (Fig. 1D,E). Our switch has louvers with a black, low IR-emissive material on the outer surface; underneath the louvers is a white, IR-emitting material. Louvers have been used to control radiative properties for spacecraft thermal management[28,29]. To achieve passive sensing and switching through opening and closing of the louvers, we use a phase-change material (PCM). These materials change from solid to liquid during heating, and have been used in building materials as a thermal capacitor to absorb large amounts of energy during temperature fluctuations and help maintain a setpoint temperature[30], and have been incorporated into radiative cooling materials[31,32]. However, these materials also change volume during phase change, which can be a downside, especially for microencapsulation[33]. In our design, we utilize this expansion, as recent work has done for solar tracking[34], to drive the configuration change of our switch.

During operation, our switch passively changes both its solar and thermal emissivities in response to its current temperature to selectively transfer heat with either the sun or deep space (Fig. 2A-D). We assume the switch's current temperature would be in equilibrium with the temperature of a building's roof into which the switch was incorporated, meaning the switch would change state based on whether the roof's temperature was above or below the setpoint, independent of the current ambient temperature. In the "black state," when the switch is below its setpoint temperature (determined by the melting point of the PCM), the PCM is solid and low-volume (Fig. 2B). This causes the louvers to lay flat on the surface, exposing the black, low IR-emissive surface, which absorbs solar radiation and emits little IR radiation to deep space. As the temperature of the switch increases beyond the setpoint, the PCM begins to melt and expand, driving a piston-linkage to open the louvers (Fig. 2C). In the "white state," when the material is above its setpoint temperature, the PCM is liquid and high-volume, and the louvers are fully open (Fig. 2D). This exposes the white, IR-emissive surface, which emits heat to deep space and absorbs little heat from the sun. To maintain a high view factor (>0.9) between the white surface and deep space, the undersides of the louvers are coated in a specularly reflecting material (See "Tile View Factor Calculation" in SI for further discussion). When the switch cools below the setpoint, the louvers close, returning the system to the black state. A simplified schematic illustrating the operation of the louvers is in Fig. SI3, and additional cross-sectional views are found in Fig. SI4.

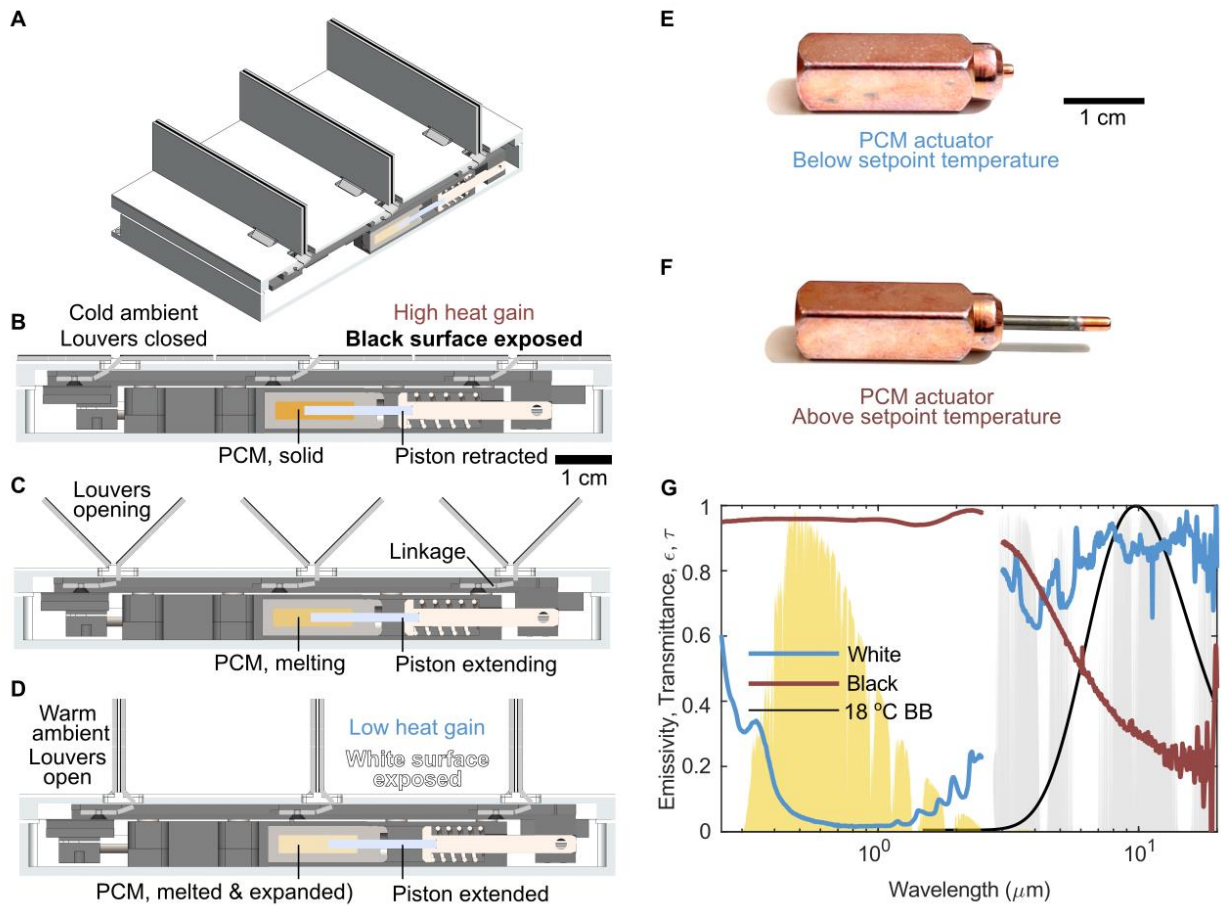

**Figure 2: The details of the design for the passive radiative switch for thermoregulation.**

(A) Isometric view with cross-section cut of the device in its open "white state."
(B) Cross-sectional view of the device in its closed "black state," with key components labeled.
(C) When the temperature rises above the set point, the PCM melts and expands, extending the piston, driving the linkage, and opening the louvers. Note that only one PCM actuator is shown in this cross section. Each PCM actuator controls three louvers (here the right-hand louvers of each pair). The second PCM actuator (not shown) controls the other three louvers.
(D) When the PCM is fully melted (<3 °C after melting begins), the louvers are fully open.
(E) Image of the PCM actuator in the retracted state. The temperature is below setpoint and the PCM is solid and low-volume.
(F) Image of the PCM actuator in the extended state. The temperature is above setpoint and the PCM is liquid and high-volume.
(G) The emissivity of the two surfaces of the adaptive switch. The black chrome-coated aluminum is shown in red, and the white BaSO₄ paint is shown in blue. Also plotted for reference are the solar emissive spectrum (yellow) and the atmospheric transmittance window (light gray), as well as the emissive spectrum of an 18 °C black body (black line).

The details of the materials that comprise the adaptive switch are as follows. On the outer surface of the louvers is a low IR-emissive, black chrome-coated aluminum (AnoBlack Cr, Anoplate Corp., Syracuse, NY). The coating is nominally 5-8 μm thick and is applied in a batch-coating process that can be applied to various metals. Black chrome coatings have been studied for decades as selective absorbers for solar heating[35] and have been shown to maintain radiative performance after prolonged non-concentrated solar exposure (1000 hr) and accelerated UV aging (1000 hr)[36]. We chose this material due to its common use. Preliminary testing shows that this coating can achieve temperatures around 100 °C in direct sunlight and reduces sub-ambient cooling at night (see Fig. SI1B). Under the louvers on the surface of the tile is the white, IR-emissive material, a barium sulfate (BaSO₄) paint. BaSO₄-based paints have been recently shown to have excellent radiative cooling performance, and the preliminary data suggests they can maintain optical performance for more than three weeks of outdoor exposure[37]. The details of both coatings are described in the Experimental Procedures section. The louvers' undersides are covered with an aluminized mylar film that mostly reflects both solar and IR radiation specularly when open. For the PCM, we used hexadecane loaded into polymer cartridges. The high latent heat of hexadecane (200 J/g) makes it a good thermal capacitor, its melting point is ideal for a temperature setpoint (~18.2 °C[38], very similar to the ideal defined in the heating/cooling degrees metric of 18.3 °C[1]), and its expansion during melting provides the actuation. We encompass a portion of the hexadecane PCM in a wax motor housing (Honda) to create a PCM actuator with the desired setpoint (Fig. 2E-F).

To characterize the emissivities of the materials that comprise the surfaces of our switch, we used Fourier-transform infrared spectroscopy (FTIR) and ultraviolet visible near infrared spectroscopy (UV-Vis-Nir) to determine absorptivity, and invoked Kirchoff's law of radiation to infer emissivity (i.e., at thermal equilibrium, spectral absorptivity is equal to spectral emissivity for the same surface) (Fig. 2G). The spectrum of the black chrome is shown in red; overall the surface has an average solar (260 nm-2500 nm) emissivity of 0.96 and atmospheric window (7-14 µm) emissivity of 0.30. The spectrum of the $BaSO_4$ paint is shown in blue; the paint has an average solar emissivity of 0.05 and atmospheric window emissivity of 0.88 (see Experimental Procedures section for calculation details). Importantly, we observe a crossover in the emissivities of the black and white surfaces, where the black surface is substantially more emissive in the solar wavelengths, but the white surface is more emissive in the thermal wavelengths. This enables our system to switch both its visible and IR properties to selectively transfer heat with either the sun or deep space.

## RESULTS

To characterize the performance of our passively adaptive radiative switch, we conducted a series of outdoor tests in which we compared the behavior of the adaptive switch in the form of a tile to non-switching control tiles, one fixed in the white state and one fixed in the black state, using the same coating as in the adaptive switch (Fig. 3A). We conducted two types of tests: temperature and power tests. For the temperature test, we measured the temperature on the back of the tiles over the course of a day (Fig. 3B). For the power test, we measured the amount of heat that needed to be added or removed to maintain the temperature on the back of the tiles within 1.5 °C of the melting point of hexadecane, 18.2 °C, (i.e., 16.7 and 19.7 °C). We refer to this 3 °C range as the deadband of the temperature controller, like found in thermostats. For both tests, we minimized the effects of parasitic conduction and convection (i.e., maximized the radiative effects) by placing the tiles in insulated boxes with a clear, polyethylene upper surface. For the power test, thermoelectric elements were used to control the temperature of the tiles and heat flux sensors were used to measure the heat fluxes and temperatures (Fig. 3D). These setups are like the ones used in a previous work[22]. The power and temperatures tests were conducted in Santa Barbara, California in July and September, respectively. Further information on the setup is detailed in the Experimental Procedures section.

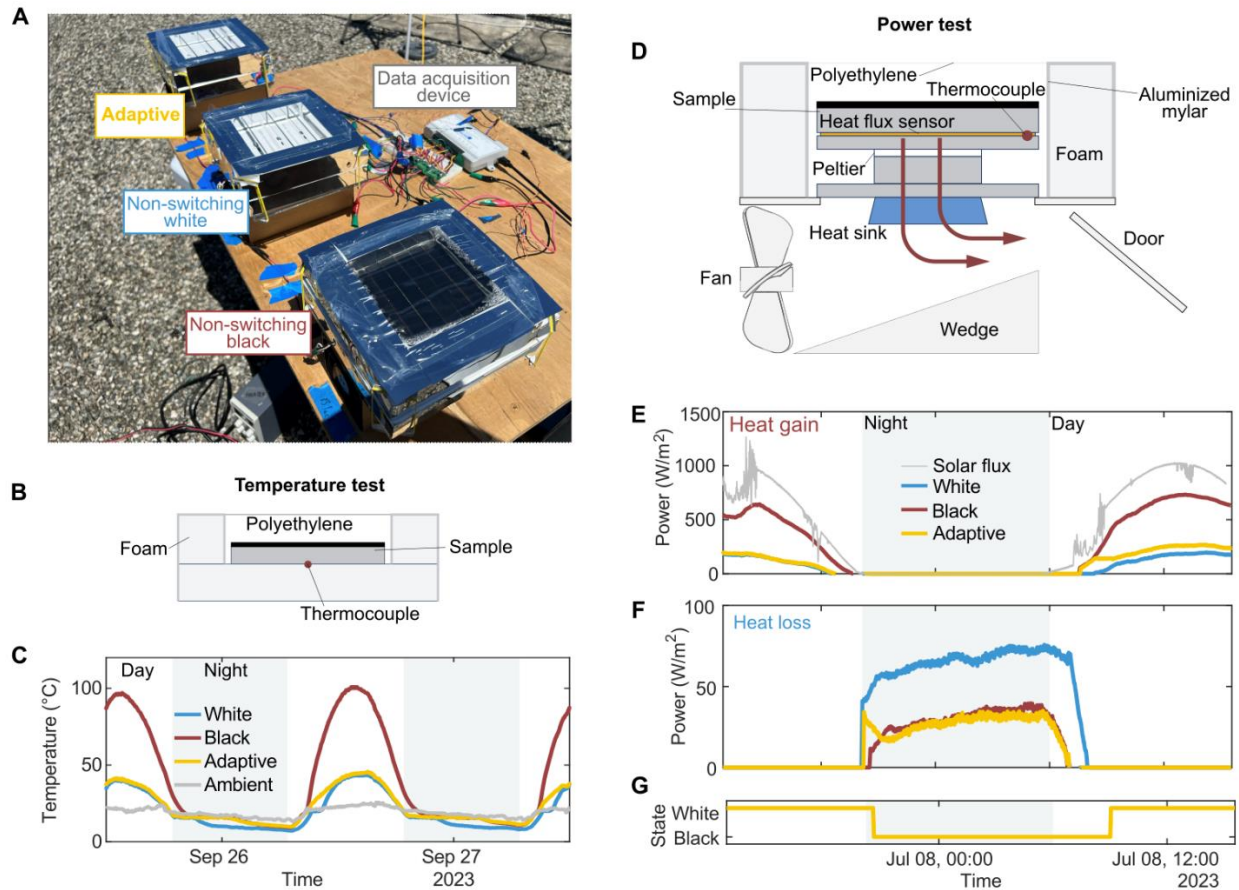

**Figure 3: Outdoor test results for the passively adaptive radiative switch show performance improvements compared to non-switching controls.**

(A) Experimental setup showing the three tested tiles (adaptive, non-switching white, and non-switching black). Each is mounted in an insulating box and outfitted with thermocouples to monitor temperature.

(B) Diagram of the setup for the first outdoor test, the temperature test. Each of the three tiles is placed in an insulated box with a thermocouple mounted below.

(C) In the temperature test, the adaptive switching tile maintains a temperature closer to the nominal setpoint (~18.3 °C) compared to the non-switching tiles. This means less active heating or cooling would be needed in a building application. During the night, the adaptive switching tile responds to its temperature being below its setpoint, by closing its louvers and maintaining a warmer temperature than the white non-switching device (minimum temperature for non-switching white: 7.0 °C; minimum temperature for adaptive switch: 9.7 °C). During the day, the adaptive switching tile opens its louvers and maintains a lower temperature than the black non-switching device (maximum temperature for non-switching black: 101.0 °C; maximum temperature for adaptive switch: 45.8 °C).

(D) Diagram of the setup for the second outdoor test, the power test. Each of the three tiles has a heat flux sensor mounted below it. Below is a Peltier element which transfers heat with a heat sink. The heat sink has forced convection cooling to help maintain it at the ambient.

(E) In the power, the temperature is maintained within a deadband (16.7 °C to 19.7 °C) while power is recorded. During the day, the adaptive switch opens, resulting in 3.1x less heat gain than the non-switching black surface.

(F) During the night, it closes, resulting in 2.6x less heat loss than the non-switching white surface.

(G) The state of the adaptive switch is shown for the power test. It switches from white to black and back to white, even though during the power test a deadband of 3 °C was enforced. This shows the small hysteresis and switching band of the adaptive switch.

The results of the temperature test show that the adaptive tile maintains a temperature closer to ideal during both the night and day compared with the black and white non-switching tiles (Fig. 3C). During the night, the adaptive tile responds to the cooler temperature, and closes its louvers. Accordingly, it behaves like the black tile and maintains a higher temperature due to reduced IR emission to the environment, closer than the white tile to the nominal comfortable temperature of 18.3 °C. The adaptive and black tiles cool, maximally, 8.6 and 9.1 °C, respectively, below nominal, whereas the white tile cools, maximally, 11.3 °C below nominal. Additionally, the lower heat loss rate prolongs the phase change of the black and adaptive tiles by about 4 hours. During daylight hours, the adaptive tile responds to the increased temperature, and the PCM expands to open the louvers. Now, the adaptive tile behaves like the white tile and maintains a lower temperature, closer to the ideal than the black tile. The adaptive and white tiles heat, maximally, 27.5 and 25.7 °C above nominal, respectively whereas the black tile heats, maximally, 82.7 °C above nominal during the experiment. The adaptive tile begins behaving like the white tile at around 9:00 when the adaptive tile's temperature track diverges from the black tile's track. See Fig. SI5,6 for temperature data with respect to nominal and ambient, respectively, and Fig. SI7 for environmental data.

The results of the second set of experiments, the heat flux test, show that the adaptive tile requires both less heat input and less heat removal during night and day, respectively, than the non-switching tiles (Fig. 3E,F). During the night, the temperature of the adaptive tile dropped to the bottom of the deadband (16.7 °C), resulting in the PCM solidifying and the louvers closing into the black state. This meant that the adaptive tile again behaved like the black non-switching tile during the night, and when compared to the white non-switching tile, reduced heat loss by a factor of 2.6. During the day, the temperature of the adaptive tile reached the top of the deadband (19.7 °C), melting the PCM and opening the louvers into the white state. Compared to the non-switching black tile, this reduced heat input by a factor of at least 3.1. We note that between 13:30 and 15:00 on July 8, the controller was unable to remove enough heat from the black tile to maintain temperature. This causes an underestimation in the needed heat removal flux and energy, suggesting that the adaptive tile was reducing heat input by an even larger factor. See Fig. SI8-11 for additional data for this experiment.

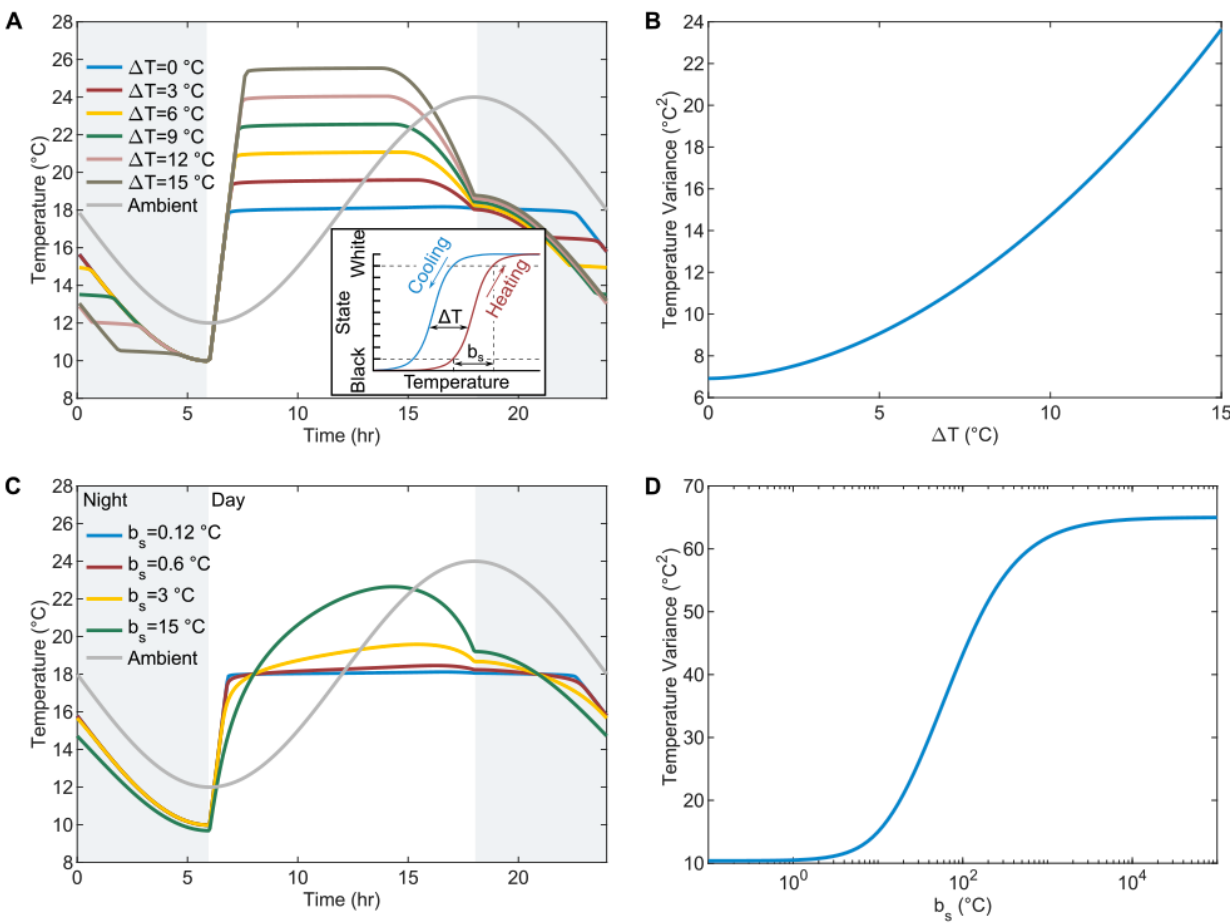

**Figure 4: Simulations show that adaptive switch performance increases as hysteresis and switching bands are minimized.**
(A) (inset) The hysteresis (ΔT) is defined as the number of degrees the temperature-state curve is shifted during cooling versus heating. The switching band is defined as the number of degrees from 10% switched to 90% switched.
(A) In the hysteresis analysis, smaller hysteresis (ΔT) results in an adaptive switch that does not overshoot the setpoint (18 °C). With increasing hysteresis, the daytime temperature increases substantially, reaching over 25 °C for a hysteresis of 15 °C. The switching band is 0.12 °C for all curves.
(B) As hysteresis increases, the temperature variance from setpoint increases super-linearly for this range of temperatures.
(C) In the switching band analysis, smaller switching bands ($b_s$) result in a more responsive behavior of the adaptive switch. As switching band increases, there is again overshoot of the setpoint with undesirably warm temperatures during the day. The hysteresis is 0 °C for all curves.
(D) As switching band increases, the temperature variance increases in a sigmoidal fashion.

Finally, we show that the adaptive switch has an important characteristic: it can switch from one state to another in a small switching band and has little hysteresis when switching back. This can be observed from the power test data, in which the device switches from white state to the black state and back again, all while the temperature is maintained within the 3 °C deadband (Fig. 3E-G).

To explore how a small hysteresis and switching band improve performance, we constructed a simulation of our adaptive switch (Fig. 4). The simulation assumes a sinusoidal temperature profile throughout the day along with a sinusoidal solar flux during daylight hours and a target setpoint temperature of 18 °C. We define hysteresis as the number of degrees that the temperature-state curve shifts between the heating direction and cooling direction; we define the switching band as the number of degrees for the temperature-state curve to change from 10% to 90% of its full value (Fig. 4a, inset). See "Hysteresis and Switching Band Model Details" in the SI for further details.

In the first simulation, we varied the amount of hysteresis, while fixing the switching band at 0.12 °C. With no hysteresis, the adaptive switch responds to the increasing temperature after sunrise (6:00) by switching states exactly at 18 °C, and maintaining this temperature throughout the day (Fig. 4a). However, when large hysteresis is added, this switching does not occur to much higher temperatures, resulting in high daytime temperatures. For instance, hysteresis of 15 °C creates a daytime temperature of over 25 °C. The variance of the switch's temperature from setpoint increases with increasing hysteresis with a super-linear trend (Fig. 4b).

In the second simulation, we varied the size of the switching band, while fixing the hysteresis to 0 °C. With a very small switching band (a very sharp transition), the adaptive switch responds again as the sun rises and the adaptive switch's temperature exceeds 18 °C. This allows it to hold its temperature at the setpoint throughout the day. In contrast, a large switching band means that the temperature will rise high above the setpoint, only decreasing once the sun becomes lower in the sky and the solar flux decreases. The variance of the switch's temperature from setpoint also increases with increasing switching band size, now showing a sigmoidal behavior (Fig. 4d).

## DISCUSSION

We demonstrated a passively adaptive radiative switch that changes state with a small switching band and low hysteresis, capable of changing from one state to the other and back within 3 °C. It can modulate the energetic cost of cooling by 3.1x and heating by 2.6x compared to non-switching tiles. Our device demonstrates the utility of small switching band and low hysteresis phase change materials for both actuation and energy storage.

Importantly, our design is highly adaptable. Different temperature set points can be achievable by simply using different phase change materials and higher radiative cooling and heating performance or high levels of durability can be achieved by different coatings. Because the coating can be applied to the rigid aluminum louvers, there are next to no constraints on materials that are compatible. This contrasts with most previous work on adaptive heating and cooling devices, which require the material to be flexible. As such, we consider our device as a platform for nearly any set of materials developed, including future materials with unprecedented performance levels.

A tradeoff that comes with this adaptability is the incorporation of moving parts. In contrast, solid-state thermal switches have an important advantage with no moving parts, and as such have garnered much attention in recent years[39]. However, to date, the level of performance of solid-state switches is far behind that of mechanical switches. For example, recent solid-state switches show the ability to switch thermal conductivity by a factor three[40] to ten[41]. This contrasts with recent mechanical switches that can achieve over 100-fold[42]. Further, another advantage of a mechanical switch design is that it allows configuration changes, such that convection properties could be switched as well (e.g., open louvers allow greater convective cooling). Future work can explore the longevity and durability of our moving-part solution, but given the advantages, we chose to use the high performance of a mechanical switch instead of the lack of moving parts of solid-state design.

Interestingly, even though our device incorporates high-performance materials, during the day, there is a larger than expected temperature increase above ideal in the temperature test and heat gain in the power test (Fig. 3C,E). We conducted simulations and experiments (see "Parasitic Heating Simulations" and "Parasitic Heating Experiments" in SI for details) to show the origins of this parasitic heating comes from the slight absorption by the underside of the louvers and the sidewalls of the test box, both of which are coated with aluminum. Further, the polyethylene convection shield results in convection cells forming which ultimately results in the heating of the tiles. During actual use, the polyethylene convection shield would not be in place, which mitigates this effect. Future studies need to be done on larger arrays to minimize boundary effects and on developing improved louver cooling methods. We also note that our tests were done in the summer, with solar flux of ~1000 W/m$^2$ and daytime relative humidity of ~60-70%, whereas most demonstrations are done in lower flux and lower humidity conditions.

Finally, an important feature of the proposed design is its simplicity and potential for low-cost production. The actuator is a mass-produced wax motor used in lawn mowers and requires no electronics or batteries and withstands thousands of cycles during regular use. The rest of the device is purely mechanical and could be made from inexpensive or recycled materials. While the presented version was designed for small-batch manufacture, the simplicity of the design lends itself easily to design for mass-manufacture.

Overall, our device advances the field of thermoregulation for buildings toward passively adaptive materials that maintain comfortable temperatures throughout the day and year. Such advances offer promise to reducing our energy consumption for heating and cooling, even as these demands continue to increase.

**EXPERIMENTAL PROCEDURES**
**Resource availability**
Lead contact: Further information and requests for resources and materials should be directed to and will be fulfilled by the lead contact, Elliot W. Hawkes (ewhawkes@ucsb.edu).

Materials availability: This study did not generate new unique materials.

Data and code availability: The data presented in this study are available from the lead contact upon reasonable request.

**Outdoor Experiments**
We conducted the power and temperature tests on the roof of Harold Frank Hall, which is located on the campus of the University of California, Santa Barbara, within 200 m of the Pacific Ocean. For both tests, we used the DAQ to record the solar flux measurement from a pyrometer (Kipp & Zonen, CMP 6). A weather station (Ambient Weather, WS-8040) recorded the air temperature, wind speed, and relative humidity along with other meteorological parameters. A camera (Go Pro, HERO9 Black) recorded both experiments. A LED flashlight provided illumination for the videos during the night. We used the videos to determine when the louvers opened and closed.

Power Tests: We ran a custom script on a DAQ (Dataq, DI-2008) to log the heat flux and T-type thermocouple readings from the heat flux sensors (FluxTeq, PHFS-09e) every 0.5s and to control the Peltier elements. An Arduino Nano microcontroller controlled the fans. We used a bang-bang (i.e., on-off) controller to maintain the temperature under the tile. We used a bang-bang controller because the DAQ only has digital outputs and many home thermostats use bang-bang control. The fans were only active when the Peltiers were active. The bang-bang control makes the heat flux and temperature data noisy. For visualization clarity, we applied a moving average filter to the heat flux and temperature data in Fig. 3 E-F. The data was smoothed over 3600 data points (30 minutes). The heating or cooling power is the heat the Peltier needs to remove or add, respectively, in order to bring the temperature to the deadband. The Peltier is active when the moving average for the Peltier state is on. The heat flux when the Peltier is inactive (i.e. temperature is within the deadband) is not plotted.

Temperature Tests: We used a K-type thermocouple (ThermoWorks, K-36X) to measure the temperature under the tiles. For the logging, we used the software (WinDaq) included with the DAQ. Data was collected every 1.0s.

**Spectral Characterization**:

UV-VIS-NIR: We used the integrating sphere accessory of the Shimadzu UV-3600 to measure the hemispherical UV-VIS-NIR reflectance spectrum of our materials. To account for the specular component for some of the materials, we used an 8° angle of incidence. Low reflectivity samples were referenced against a 10% diffuse reflectance standard (Labsphere, Speclatron) and high reflectivity samples were referenced against a 99% diffuse reflectance standard (Labsphere, Speclatron).

FTIR: We used an integrating sphere with a Bruker Invenio-R FTIR spectrometer to measure the hemispherical infrared reflectance spectrum of our materials. Diffuse gold was used as the reference. The materials measured are opaque, so emissivity, $\epsilon$, is related to reflectance, $\rho$, by $\epsilon = 1 - \rho$.

The solar emittance is the ASTM 1.5 solar spectrum weighted average. It has the form:

$$\epsilon_{solar} = \frac{\int_{260\,nm}^{2500\,nm} \epsilon(\lambda) I_{ASTM1.5}(\lambda) d\lambda}{\int_{260\,nm}^{2500\,nm} I_{ASTM1.5}(\lambda) d\lambda}$$

$I_{ASTM1.5}$ is the ASTM 1.5 solar intensity spectrum and $\lambda$ is the wavelength.

The atmospheric emittance is a simple average. It has the form:

$$\epsilon_{7-14\mu m} = \frac{\int_{7\mu m}^{14\mu m} \epsilon(\lambda) d\lambda}{\int_{7\mu m}^{14\mu m} d\lambda}$$

**Black and White Coating Preparation:**

Details of black coating process: Prior to coating, we manually polished 0.006" thick 6061-O aluminum foil (ESPI Metals) with aluminum polish (Mothers). After cleaning, the foils were sent to Anoplate Corp., Syracuse, NY for the black chrome coating. Anoplate performed their commercial coating process to create the coating they call "AnoBlack Cr." After coating, the "knap" on the surface of the coating was removed, because it sometimes left a powdery residue. The coating was removed by rubbing the surface with an acetone-soaked cotton swab with light pressure. Removing the "knap" slightly reduced solar absorbance and significantly reduced thermal emittance (see Fig. SI2).

Details of white coating process: We adapted the ultra-white coating by Li et al.[36]. We mixed our paint in small batches. Each batch of paint contains the following:
- 4.28 g $BaSO_4$ (Carolina Biological Supply)
- 0.30 g acrylic emulsion (Dow, Rhoplex AC-2235M)
- 4.12 g DI water

DI Water is added to the $BaSO_4$ and the mixture is mixed until smooth and homogenous on a stir plate. The acrylic emulsion is then added and mixed until fully incorporated (~2 minutes). Prior to coating, the tile surface was primed with a self-etching primer (Rustoleum). The coating was applied to the tile with a drawdown bar. Multiple coatings were applied until the nominal thickness of 0.5 mm was achieved.

**Details of Device Design:**
We fabricated the tiles primarily with a mix of machined aluminum parts and 3D printed parts.

The tile bodies are made with 6061 aluminum and machined with a CNC milling machine.

The louvers and linkages are made of laser cut 5052 aluminum. The hinges of the linkages are made of polyester fabric which is adhered to the aluminum with cyanoacrylate adhesive (Henkel, Loctite 401) and spectra thread. The aluminized mylar film was adhered to the louver with transfer tape (3M). Neodymium magnets (totalElement) are pressed into the louvers. These magnets help with thermal contact in the black state and keep the flaps open in the white state. See Fig. SI4D for cross sectional views.

The hinges connecting the louvers to the top surface of the tile are made of polyester fabric. The fabric is adhered to the louvers with transfer tape (3M) and the hinges are sewn into the tile with spectra fiber (PowerPro). The selective absorbers are adhered to the top of the fabric with transfer tape (3M).

The hexadecane wax is contained in clear 3D printed cases (FormLabs, Clear Resin). The other 3D printed parts (shown in dark gray in Fig. 2A-D) are made with nylon with chopped carbon fiber (MarkForged, Onyx).

The different components are connected to each other with screws (McMaster). We filled gaps with boron nitride heat transfer paste (Miller-Stephenson, ReleaSys HT Paste).

**Acknowledgements.**
We thank Chenxi Sui, Isabella Ramirez, Ruotong Zhang, Sicheng Wang, Harnoor Lal, and Connie Berdan for their prototyping, fabrication, and experimental setup assistance across the different iterations of this project. We also thank Aryan Zaveri of the Raman Lab at the University of California, Los Angeles for performing the FTIR measurements and Michael Gordon of the University of California, Santa Barbara for help with early FTIR measurements. We also thank the UCSB TEMPO facility at the MRL for help with the UV-VIS-NIR Measurements. Charles Xiao is supported by the NSF Graduate Research Fellowship Program (2139319), and the work is partially supported by a UCSB CNSI Challenge Grant.

**Supplementary Information**

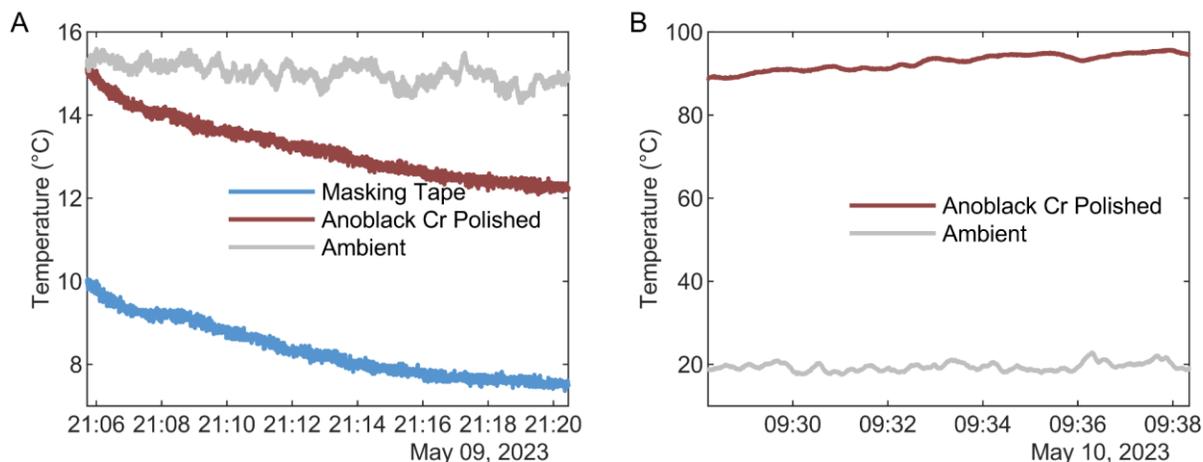

**Figure SI1: Black material characterization.**
(A) Temperature trace of the black material during a brief nighttime outdoor test. The material was compared to a piece of aluminum shim of similar thickness covered in masking tape, which is highly emissive. During the test, the black material cooled about 3°C below ambient whereas the emissive sample cooled over 7°C below ambient.
(B) Temperature trace of the black material during a brief daytime outdoor test. Even well before solar noon, the material reaches temperatures approaching 100°C. For these tests, the materials were placed in an insulated foam box with a polyethylene convection shield. Ambient temperature was estimated by sticking a thermocouple into the air. There was sufficient wind during the daytime test to limit radiative heating of the probe.

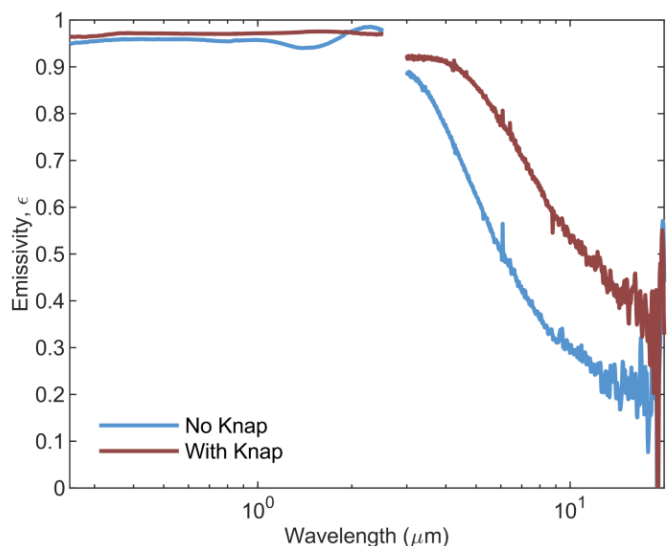

**Figure SI2: Black material, effect of knap.**
Emissivity spectrum of the black material with and without the "knap." Removing the knap reduces the solar emissivity slightly and significantly reduces thermal emissivity.

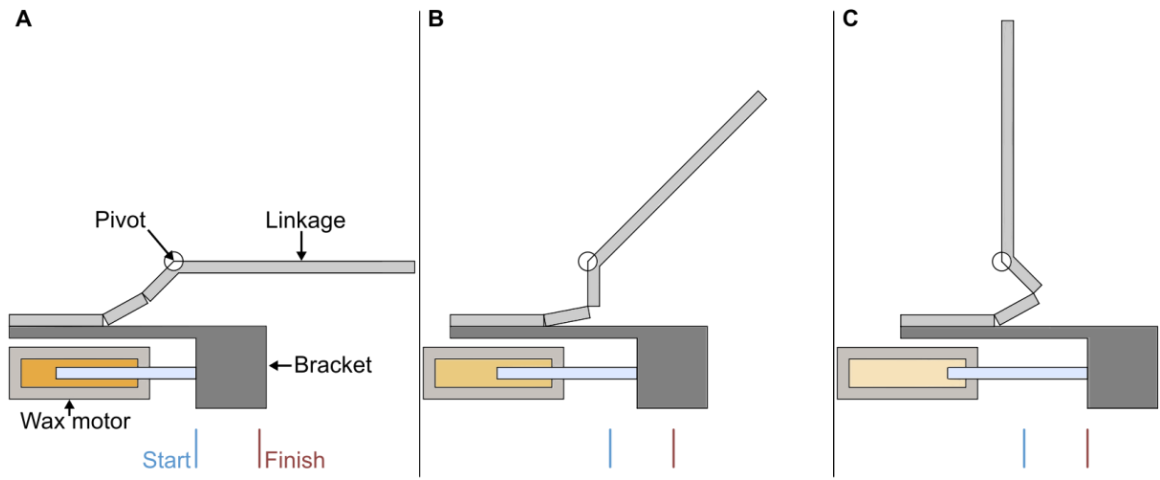

**Figure SI3: Schematic of the louver linkage**
(A) Device in the "black state," with the louver closed.
(B) During heating, the wax melts, extending the piston out of the motor, and pushing the bracket. The bracket in turn pulls the linkage, which begins to lift the louver, which rotates about the pivot point.
(C) Once fully heated, the wax is fully expanded, and the piston is fully extended, the bracket is at the end of its travel, and the linkage forces the louver into a vertical position.

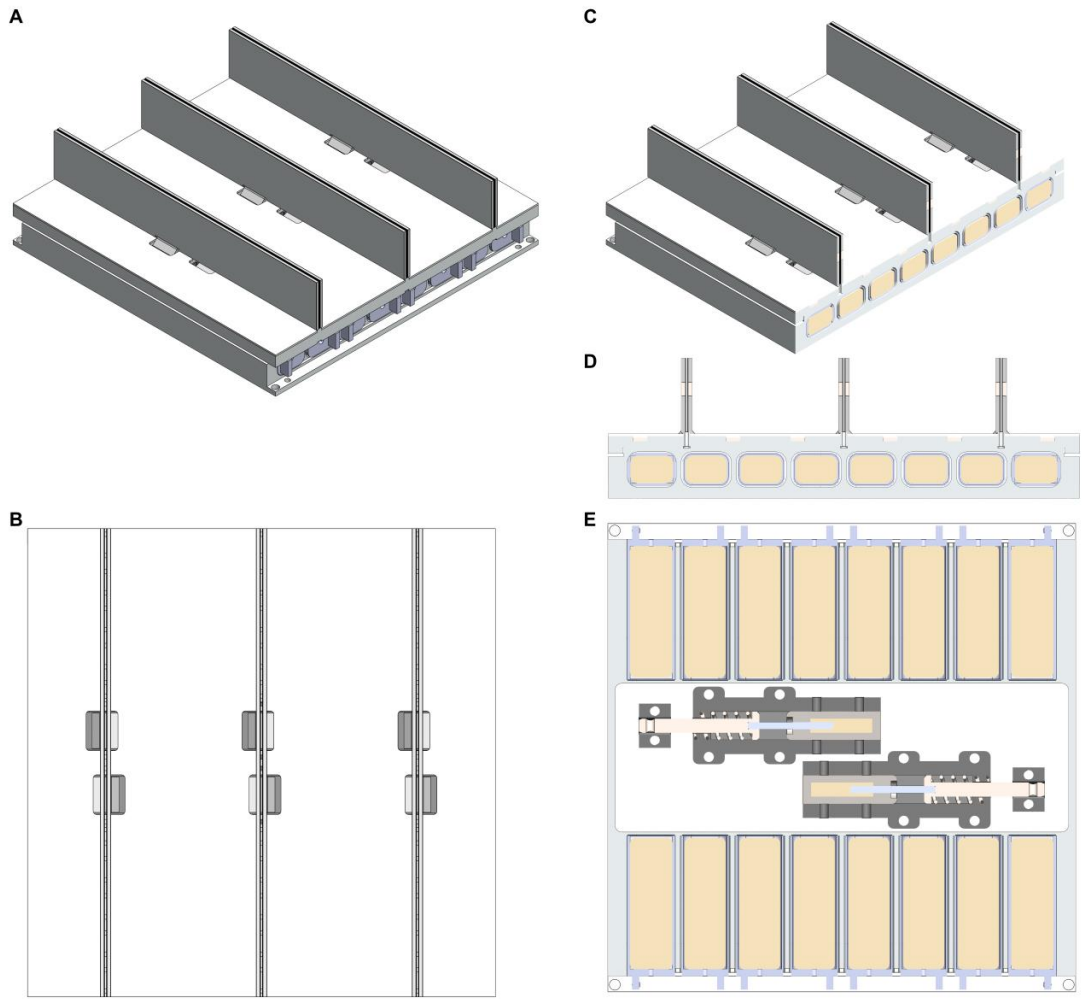

**Figure SI4: Additional views of passively adaptive switch.**
(A) Isometric view of the full device with louvers open (white state).
(B) Top view of full device with louvers open (white state).
(C) Isometric view with cross-sectional cut through the PCM tanks. Note that these tanks are separate from the smaller PCM actuators.
(D) Cross-sectional side view through the PCM tanks.
(E) Top view of slice through the PCM tanks and PCM actuators (two PCM actuators, facing opposite directions). Each PCM actuator controls three louvers through its linkage.

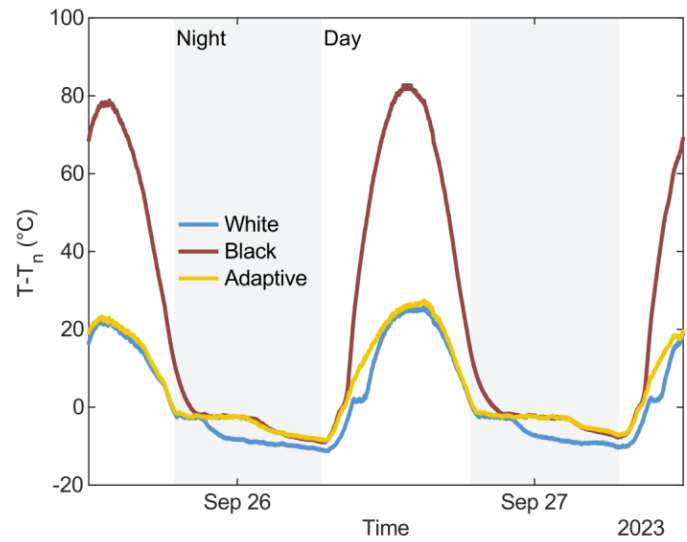

**Figure SI5: Temperature test with respect to a nominal comfortable temperature of 18.3 °C.**
Plot of the difference between the tiles' temperatures and the nominal comfortable temperature of 18.3 °C.

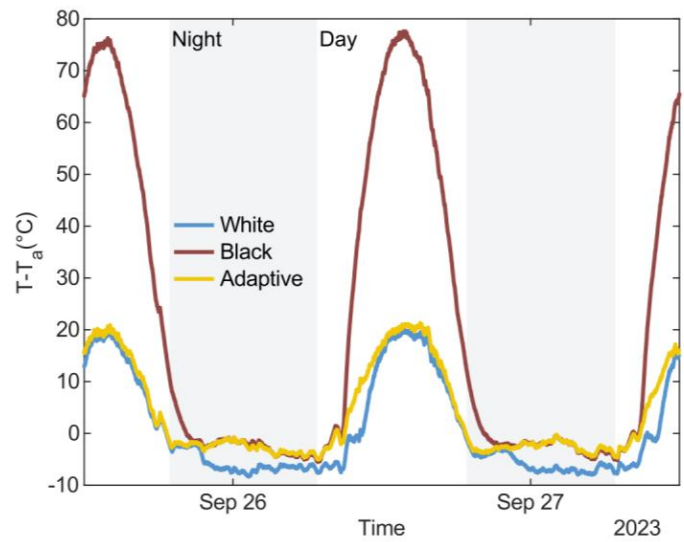

**Figure SI6: Temperature test with respect to ambient.**
Plot of the difference between the tiles' temperatures and the ambient air temperature

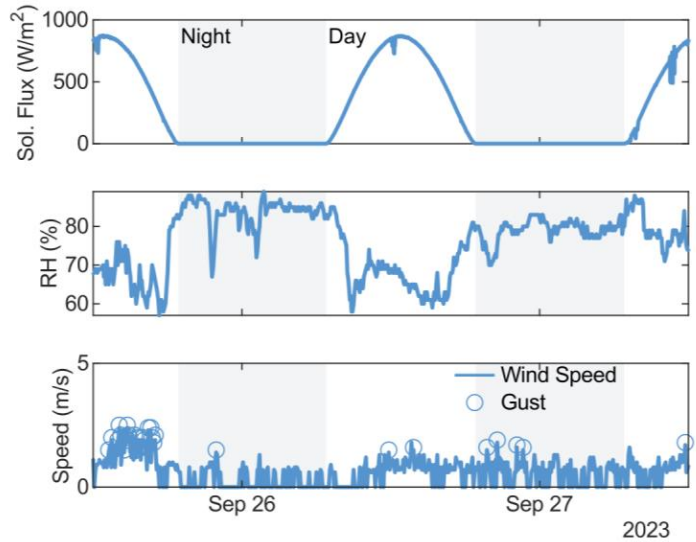

**Figure SI7: Temperature test environmental data.**
Plot of the solar flux, relative humidity (RH), and wind speed for the temperature test.

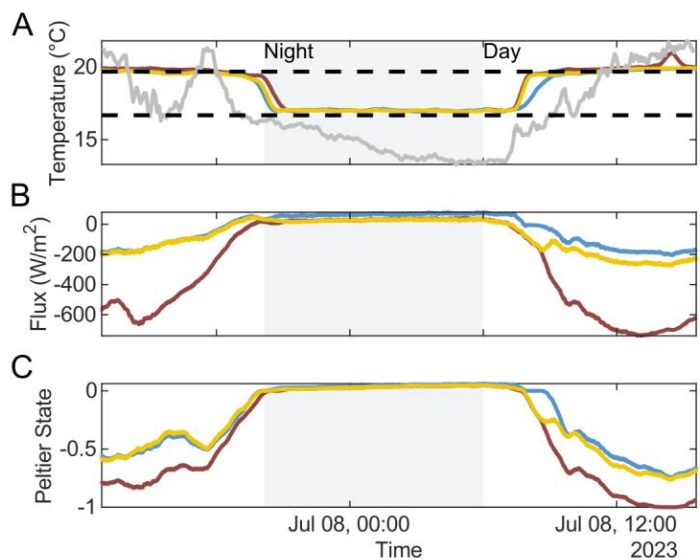

**Figure SI8: Power test additional data.**
(A) Plot of the tiles' temperatures during the power test. The red, blue, and yellow curves are the black, white, and adaptive tiles' traces, respectively. The gray curve is the ambient temperature. The dashed black lines show the deadband.
(B) Plot of the heat flux across the heat flux sensor.
(C) Plot of the average state of the Peltier element. A state of -1 represents the Peltier removing heat from the tiles and a state of 1 represents the Peltier adding heat to the tiles. A state of 0 represents the off state of the Peltier.

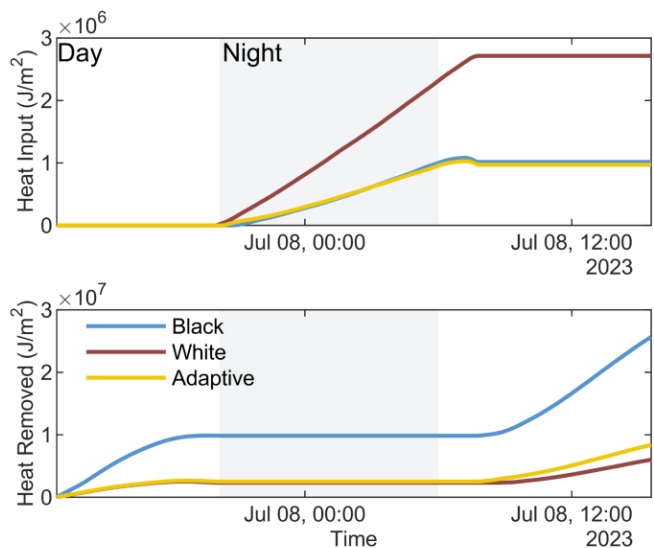

**Figure SI9: Power test total energy.**
Plot of the energy inputted and removed from the tiles by the Peltier. The adaptive tile requires 2.6x total heating than a tile fixed in the white state when measured from the maximum points. It requires less 3.1x less heat removal than a tile fixed in the black state when measured from the maximum point. The decrease in heat input is an artifact of data smoothing.

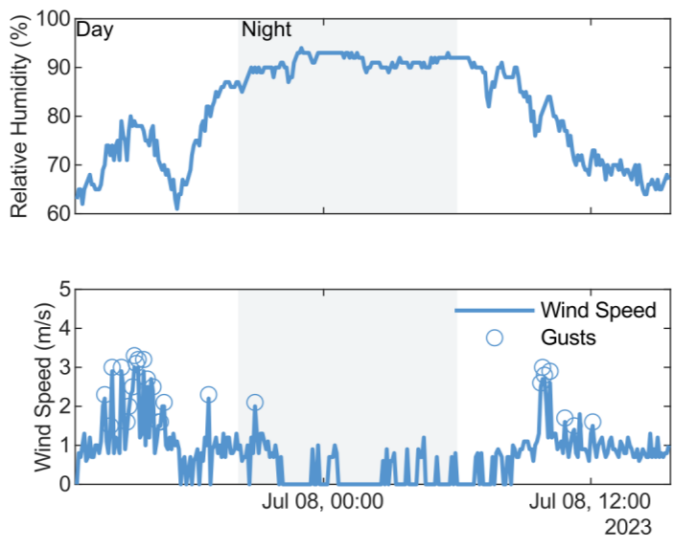

**Figure SI10: Power test environmental data.**
Relative humidity and wind speed plot for the power test

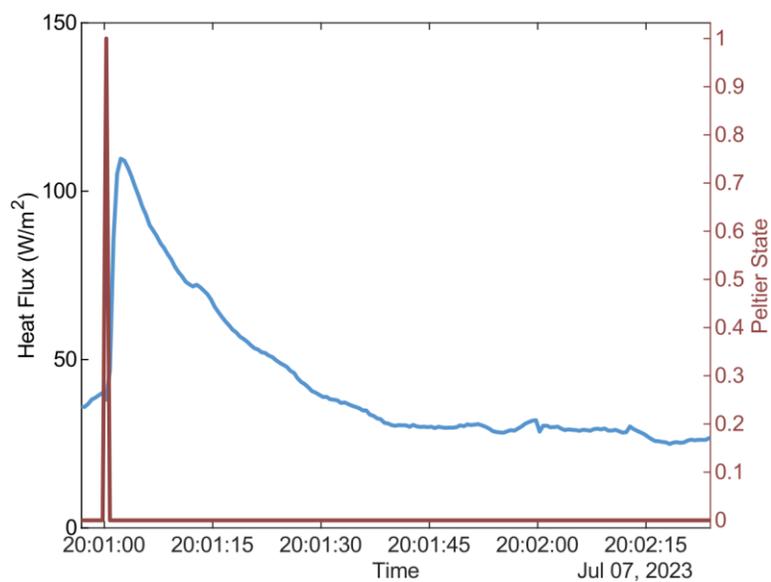

**Figure SI11: Peltier response.**
We estimate that each pulse of the Peltier affects heat flux readings for about 1 minute. This was measured on the tile fixed in the white state.

**Tile View Factor Calculation**

We derive and calculate the specular view factor from the white surface to the sky to show that the louver structure does not significantly limit the view factor of the tiles to the sky.

The geometry of the tile can be simplified and idealized as 4 infinitely long flat plates arranged as a trapezoid. The base is the white surface, the sides are the underside of the louvers, and the top is the sky (Fig. SI12A). We assume that the white surface and sky have no diffuse reflectance component.

With perfectly specularly reflecting side walls (i.e., underside of the louvers), the view factor between the sky and the white surface can be easily calculated by invoking the specular reciprocal relationship[1]

$$A_i F_{ij}^s = A_j F_{ji}^s$$

$A$ is the area of the surface $i$ and $F_{ij}^s$ is the specular view factor between surfaces $i$ and $j$.

For angles less than vertical ($\theta < 90°$), the angle subtended between the sky and the walls is $> 90°$; thus, all the radiation emitted from the sky reaches the white surface either directly or through specular reflection. The specular view factor between the sky and the white surface, $F_{sw}$, is 1. The area of the sky, $A_s$, is proportional to $1 - \cos(\theta)$. Therefore, the view factor between the white surface and the sky is:

$$F_{ws}^s = 1 - \cos(\theta), \quad 0 \leq \theta \leq \pi/2$$

If the walls have no specular component, then the view factor between the white surface and the sky can be calculated by the view factor relationship for infinite parallel plates with vertically aligned midlines[2]. Thus, the view factor is:

$$F_{ws} = \frac{1}{2}\left(-\sqrt{\csc(\theta)^2} + \sqrt{\csc(\theta)(-4\cot(\theta) + 5\csc(\theta))}\right)\sin(\theta)$$

For cases where there are both diffuse and specular components, we use a ray tracing simulation (COMSOL) to compute the view factor. Fig. SI12B plots the view factor as the wall diffuse reflectance and angle is varied.

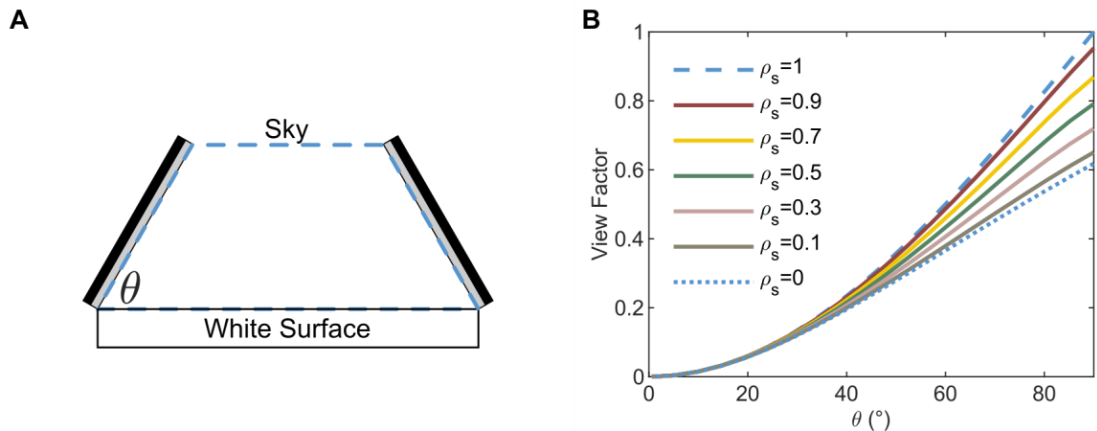

**Figure SI12: White Surface to sky view factor.**
(A) Idealized geometry of the tile. The trapezoidal geometry is drawn in the dashed blue line.
(B) Plot of the view factor from the white surface to the sky for varying louver angles and specular reflectances. Analytical solutions are plotted in the dotted and dashed lines. The simulated solutions for those reflectances lie directly on the analytical solution curves.

Fig. SI12B shows that highly specular materials, such as polished aluminum, enhances radiative transfer from the white surface to the sky.

**Parasitic Heating Experiments**

To explain the higher than expected temperatures of the tiles in the white state, we conducted a series of experiments and simulations. We hypothesize that the formation of convection cells within the insulated boxes leads to heating of the tiles. Our switch is designed such that when the louvers open, the primary pathway of heat flow from the louvers to the surface is only through the high-thermal-resistance hinges. This assumes that the louvers are exposed to the ambient air, and heat is thus primarily convected away into the surrounding air. However, during testing, the tiles are covered by a polyethylene convection shield, which largely prevents convection with the ambient. The louvers heat up in direct sunlight (they are coated with aluminum), and there is possibly a convection cell established underneath the convection shield, transferring heat from the louvers to the surface. Additionally, the internal sidewalls of the insulated box are also aluminum, creating four more surfaces that can heat (which are not present for shallow samples without louvers).

We measured the temperature of the louvers by attaching a K-type thermocouple (Thermoworks, K-36X) to the surface of the aluminized mylar on the sun facing side (Fig. SI13A). The thermocouples were then covered with aluminized mylar to better match the radiative loading conditions that the louvers see. We measured the temperatures for cases with and without the polyethylene convective shield. For the case with the convective shield, it was done on the tile that is fixed in the white state. For the case without the convective shield, it was done on the adaptive tile in the white state.

The local air temperature was measured to between 24-26 °C.

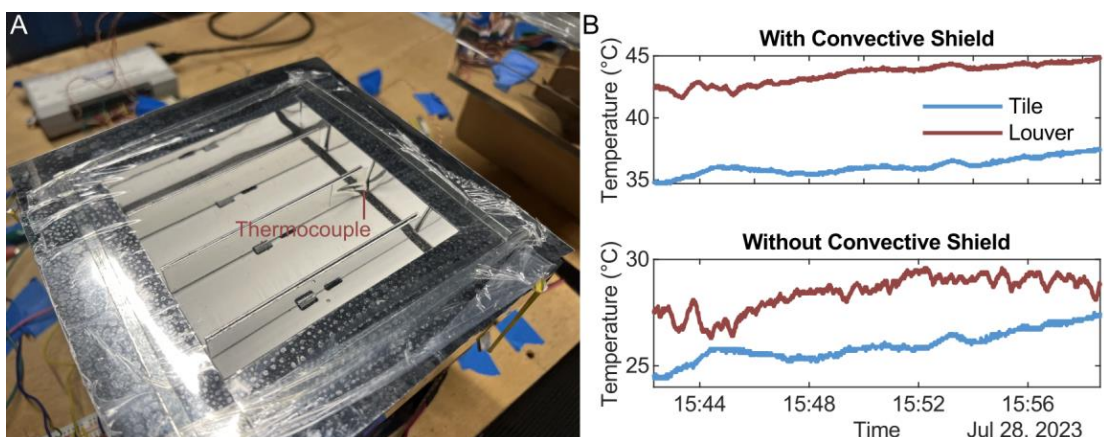

**Figure SI13: Louver tests.**
(A) Picture showing thermocouple placement.
(B) Temperature traces for tiles and louvers with and without the polyethylene convective shield.

As predicted (shown in Fig. SI13B), the louver temperature is higher than that of the tiles. This might result in a convection cell forming within the enclosure. Hot air rises from the louvers and then cools and sinks from the cool convection shield surface. Removing the convection shield lowers the temperature of the louvers and tiles. This suggests that in practical applications, where there are no convection shields, the heating is minimized.

For the second series of tests, we investigated whether changing the underside of the louvers to white paint improved performance and the influence of the aluminized sidewalls of the insulated box for the case with the convective shield. For the former, we used magnets to attach white painted shims to the underside of the louvers of the tile fixed in the white state (Fig. SI14A). We then compared the temperature evolution of that tile to the adaptive tile in the white state with the original aluminized louver underside. For the latter, we painted two aluminum plates with the same planar dimensions as the tiles white and recessed them to different depths. At the deep depth, the plate's surface sat about 19 mm below the convection shield, a level like that of the white surfaces of the tiles. At the shallow depth, the plate's surface sat about 5 mm below the convection shield.

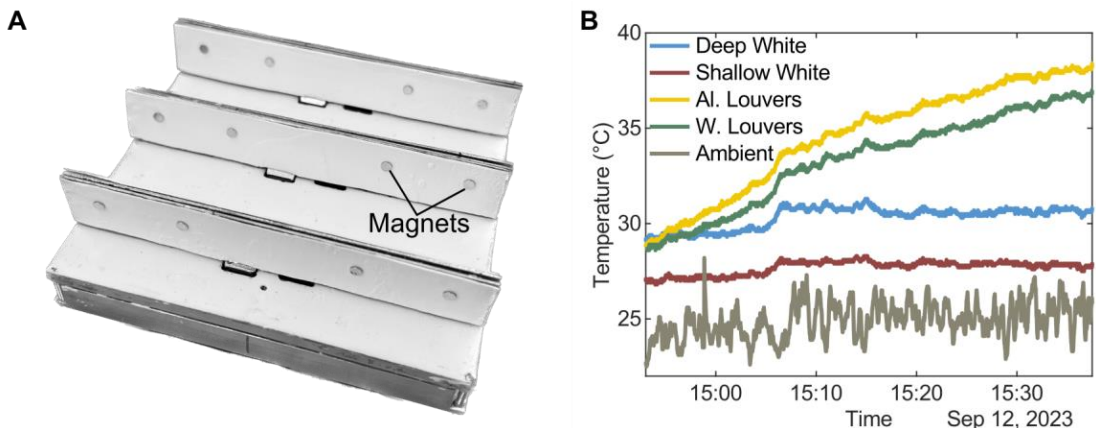

**Figure SI14: Second series of tests**
(A) Tile with white louver undersides.
(B) Temperature traces for tiles with aluminized (Al. Louvers) and white (W. Louvers) louver undersides, for white painted plates at deep (Deep White) and shallow (Shallow White) depths, and the ambient air temperature. The ambient air temperature was measured by a shaded K-type thermocouple with good wind flow over it.

Fig. SI14B plots the temperature evolution of the tiles and plates. The tile with the white louver underside is cooler than the tile with the aluminized undersides by about 1.5 °C. This can be potentially explained by the cooler louver temperatures (due to the higher thermal emissivity), which ultimately results in less heat transferred from the louvers to the tile. The plate at the deep depth is about 2.5 °C hotter than the plate at the shallow depth and more than 6 °C cooler than the tiles with louvers. At equilibrium, the temperature difference between the tiles with louvers and flat plates is expected to be even greater. This suggests that convective heat transfer from the sidewalls and the louvers can be significant. Theoretically, the highly specular side walls minimally impede radiative transfer from the surface to the sky and radiates minimally to the surface.

For the third series of tests, we repeated the second series of tests, but removed the convection shield and measured the temperature of a tile fixed in the black state. Fig. SI15 plots the temperature evolution for this test. Increased convection eliminated the performance difference between tiles with white and aluminized louver undersides and reduced the temperature difference between the deep and shallow white plates to less than 1.5 °C. The tiles with the louvers are about 4.5 °C hotter than the deep white plate and about 25 °C cooler than the black tile. These results suggest that in real applications (in which a convective shield would not be used), the louver concept will result in some small amount of heating in sunlight but is still an effective way to modulate building heating and cooling.

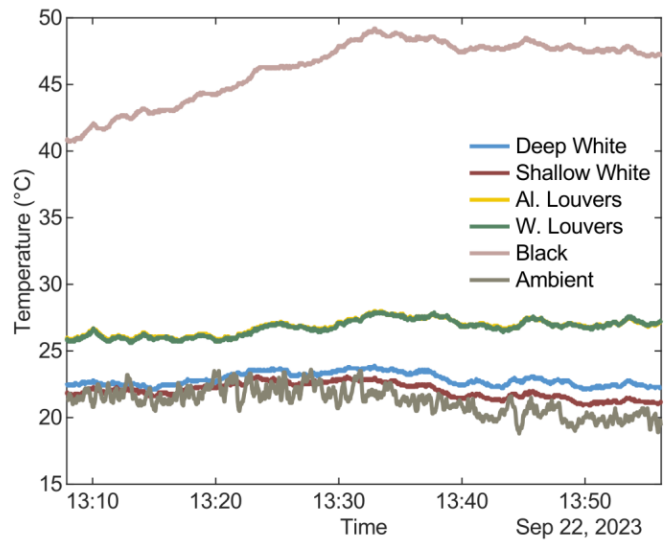

**Figure SI15: Temperature evolution for experiment in which the convective shield was removed.**
In this case, the louvers result in a much smaller amount of heating than with the convective shield.

**Parasitic Heating Simulations**

To better elucidate the origins of the heating of the surface in the white state, we conducted several simulations using COMSOL (version 5.6). While these results do not perfectly match experiments, they nonetheless provide insight into the problem. These studies show that some of the observed heating comes from radiative heat transfer from the louvers, but it primarily comes from heat conduction and convection between the louvers and the plate and from boundary effects.

In this section, we first detail the setup of the models and then report on the results of the studies. Our studies are divided into two parts. For simplicity, the first study has no fluid. Heat is transferred without convection. The second part adds a laminar flow model to model the convection effects.

*Setup*

Geometry: We used simplified geometries for our studies. Fig. SI16A represents a typical section of the tile. The two tall thin rectangles represent the louvers, and the bottom rectangle represents the plate. For simplicity, they are made of aluminum. There are three variants of the geometry analyzed: flat, normal, and edge. The flat geometry has no louvers, the normal geometry is a typical section of tile, and the edge geometry roughly describes the geometry of the sections along the sides of the insulated boxes used for testing. Fig. SI16B shows the geometry used to analyze the convection occurring inside the insulated boxes. $t_g$ is the size of the air gap between the polyethylene film and the top of the louvers. For our studies, $t_g$ is 0.03175 m. Table SI1 summarizes the parameters used for the study.

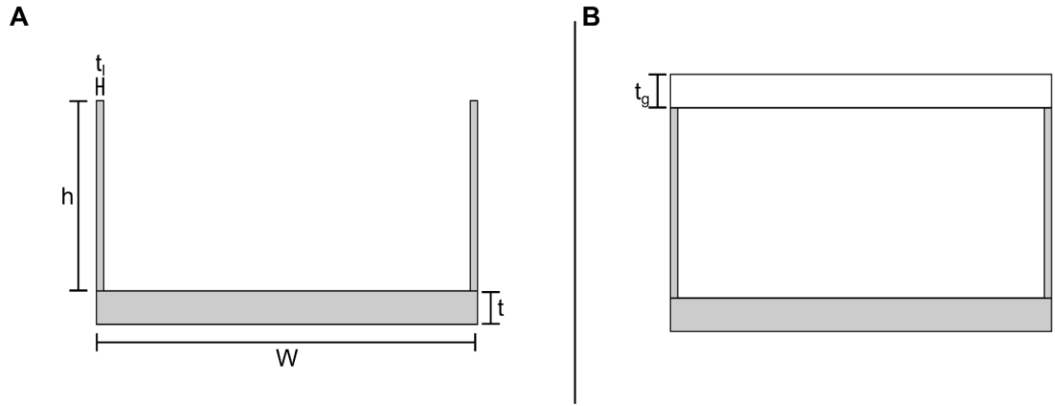

**Figure SI16: Geometries analyzed**
(A) Simplified geometry of a section of tile. Parameter values are given in Table SI1.
(B) Geometry used for the studies with convection. The white rectangles represent the air.

| Table SI1: Simulation geometry parameters | | | | |
|---|---|---|---|---|
| Geometry | W (m) | t (m) | h (m) | $t_l$ (m) |
| Flat | 0.034925 | 0.003175 | N/A | N/A |
| Normal | 0.034925 | 0.003175 | 0.0174625 | 7.9375E-4 |
| Edge | 0.0174625 | 0.003175 | 0.0174625 | 7.9375E-4 |

Radiative model: For the radiative heat transfer, we use ray tracing, because of the possibility of specular reflection. The edges marked in red and blue of Fig. SI17A participate in the radiative heat transfer. For the tests with convection, the air is assumed to be non-participatory. The red lines in Fig. SI17A mark the undersides of the louvers; the blue marks the white surface. For our studies, we compare the impacts of white and aluminized louver undersides (the red areas). Table SI2 gives the radiative properties of the white and aluminized conditions. For our ambient model, we assume that the ambient temperature is 293.15 K. The ambient atmosphere is divided into 9 bands. The simplified 9 band model is based on the MODTRAN3 atmospheric emittance model for mid-latitude summer. Fig. SI17B compares the two models. To model the effects, the sun is modeled as a blackbody at 5780 K and radiates at an intensity of 1000 W/m$^2$. We use two Sun elevations of 90 and 54.22° (corresponding to the maximum sun elevation on September 26, 2023) for our studies.

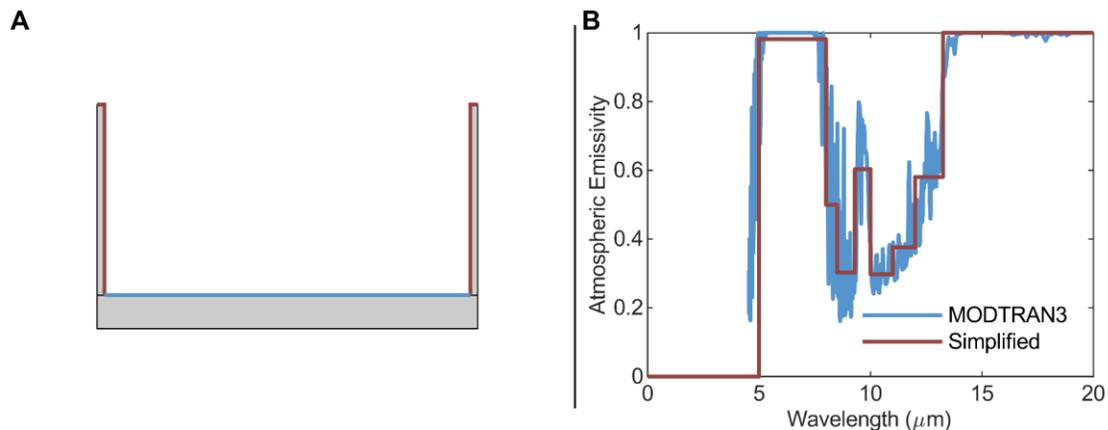

**Figure SI17: Radiative models**

(A) Radiative heat transfer occurs on the lines marked in red and blue. The red and blue line represents the louver undersides and the white plate surface, respectively.
(B) Comparison between the MODTRAN3 emittance model and the simplified 9 band model

| Table SI2: Radiative properties of the materials | | | | |
|---|---|---|---|---|
| Material | $\epsilon_{0-5\mu m}$ | $\rho_{s,0-5\,\mu m}$ | $\epsilon_{5-\infty\mu m}$ | $\rho_{s,5-\infty\,\mu m}$ |
| White | 0.05 | 0 | 0.88 | 0 |
| Aluminum (Al.)[3] | 0.089 | 0.911 | .074 | 0.926 |

Fluid Model: A compressible low Mach number laminar flow model is used to model the fluid flow.

Conductive and Fluid Boundary Conditions: To simulate a large tiling, we use a periodic boundary condition for the conductive and fluid models. These are marked in red on Fig. SI18 A and B. The underside of the tile is assumed to be thermally isolated. For the cases where there is poor thermal conduction between the louvers and the plate, we assume that the boundary is insulated. For cases where there is good thermal contact, the boundary is not insulated. In Table SI3, the Insulation column states whether this boundary is insulated or not. The insulated boundary conditions are marked in blue in Fig. SI18A. The surfaces for the fluid model are non-slip. This is marked in green on Fig. SI18B. To model the convection, the heat transfer into the ambient at the interface of the polyethylene film, we apply a convective heat flux condition at the boundary marked in yellow of Fig. SI18B. We use a convection coefficient of 10 W/m²K.

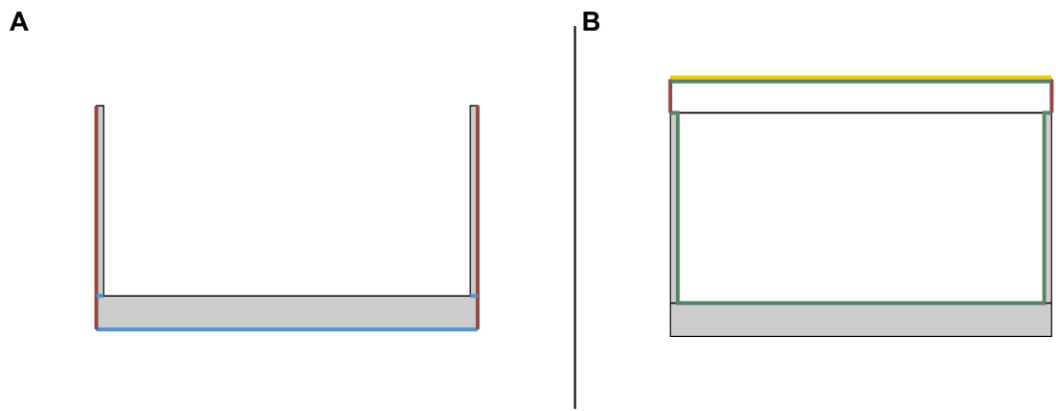

**Figure SI18: Boundary conditions**
(A) Boundary conditions used for the models without fluid flow.
(B) Boundary conditions used for the model with fluid flow. The boundary conditions in A also apply.

| Table SI3: Summary of no convection results | | | | | | |
|---|---|---|---|---|---|---|
| Test | Geometry | Louver | Sun El. (°) | Insulation | $T_{plate}$ (K) | $T_{louver}$ (K) |
| 1 | Flat | N/A | 90 | N/A | 290.00 | N/A |
| 2 | Flat | N/A | 54.22 | N/A | 287.81 | N/A |
| 3 | Normal | Al. | 90 | Yes | 293.94 | 360.61 |
| 4 | Normal | Al. | 90 | No | 298.86 | 298.89 |

| 5 | Normal | Al. | 54.22 | Yes | 292.84 | 379.70 |
|---|---|---|---|---|---|---|
| 6 | Normal | Al. | 54.22 | No | 299.92 | 299.96 |
| 7 | Normal | White | 90 | Yes | 298.36 | 295.19 |
| 8 | Normal | White | 90 | No | 296.99 | 296.98 |
| 9 | Normal | White | 54.22 | Yes | 293.43 | 295.32 |
| 10 | Normal | White | 54.22 | No | 294.26 | 294.26 |
| 11 | Edge | Al. | 90 | Yes | 297.77 | 361.43 |
| 12 | Edge | Al. | 90 | No | 306.64 | 306.67 |
| 13 | Edge | Al. | 54.22 | Yes | 297.73 | 381.05 |
| 14 | Edge | Al. | 54.22 | No | 310.40 | 310.43 |

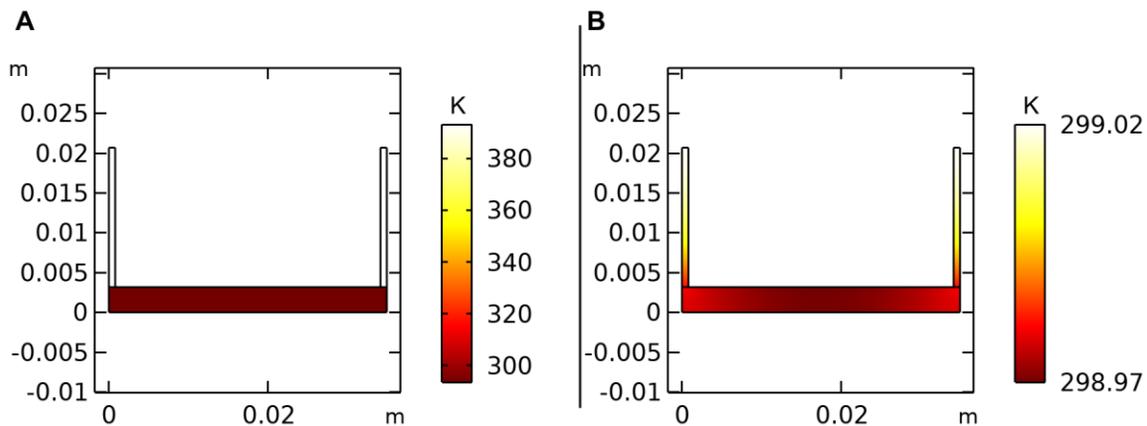

**Figure SI19: Temperature profiles for the no convection tests.**
(A) Test 3 temperature profile. Other tests with insulated louvers look similar but with different temperatures.
(B) Test 4 temperature profile. Other tests with un-insulated louvers look similar but with different temperatures.

Results

Table SI3 summarizes our simulation results for the cases with no convection as the geometry, louver underside material, Sun elevation angle, and louver-plate insulation are varied. Fig. SI19A plots the temperature profiles for tests of insulated and uninsulated louver plate contact. The flat plate geometry results in the lowest equilibrium temperature and is the only configuration that achieves radiative cooling. All tested configurations with louvers result in heating above the ambient. This is explained by the radiative heating from the louvers onto the plate. The aluminized louvers reach high temperatures, and the white painted louvers are intrinsically emissive. For cases where the louvers are well insulated from the plate, aluminized louvers result in lower plate temperatures than white ones because of the better sky view factor. However, in cases where the louvers and plates are in conductive contact, white louver undersides result in lower plate temperatures. The white louvers have lower thermal loading from the sun and the plate, and thus, deliver less heat to the plate when they are in conductive contact. Additionally, the white louvers are less affected by the Sun elevation angle; the high emissivity dissipates the increased solar loading from lower Sun elevations (until shading becomes significant). In contrast, the aluminized louvers cannot as efficiently dissipate the increased loading. This results in higher louver temperatures when they are thermally

isolated from the plate and higher plate temperatures when they are in good contact with the plate despite the lower direct solar loading on the plate.

For the tested conditions, the highest plate temperatures occur at a Sun elevation of 54.22° for the edge geometry with aluminized louvers in conductive contact with the plate. Remarkably, the temperatures are somewhat close to the measured temperatures during the temperature tests. This suggests that edge effects from the insulated box are a major contributor to the tile heating.

Additionally, we simulated the case where the louver surfaces were fixed to the ambient temperature (Table SI4). This is a rough approximation for when there is enough convection to cool the louvers. In this case, systems with aluminized louvers are only slightly warmer than the flat plate. Further, they are cooler than systems with white louvers due to the improved view factor to the sky. This shows that high performance can be achieved with aluminized louvers if louver heating is minimized.

| Table SI4: Summary of Louver Fixed at Ambient Temperature Tests | | | | | | |
|---|---|---|---|---|---|---|
| Test | Geometry | Louver | Sun El. (°) | Insulation | $T_{plate}$ (K) | $T_{louver}$ (K) |
| 15 | Normal | Al. | 90 | Yes | 290.54 | N/A |
| 16 | Normal | Al. | 54.22 | Yes | 288.04 | N/A |
| 17 | Normal | White | 90 | Yes | 297.64 | N/A |
| 18 | Normal | White | 54.22 | Yes | 292.63 | N/A |

Next, we couple a simple laminar flow model to the simulation. For simplicity, we build this model upon the conditions of Test 3 (Table SI3). Fig. SI20A and B plot the resulting temperature and velocity profiles at a time of 240 minutes. At that time, the temperature and flows have reached steady state. The resulting convection cell (Fig. SI20B) has the effect of homogenizing temperature even if there is no conduction between the louvers and the plate. The equilibrium plate and louver temperatures are 296.63 and 296.65 K.

Additionally, we analyze the effects of a taller air gap, a $t_g$ of .01905 m (Fig. SI21). This shows, unsurprisingly, that hot air rises from the louvers. If there was air flow, (e.g., from wind), then the hot air would be blown away, reducing heating on the tiles.

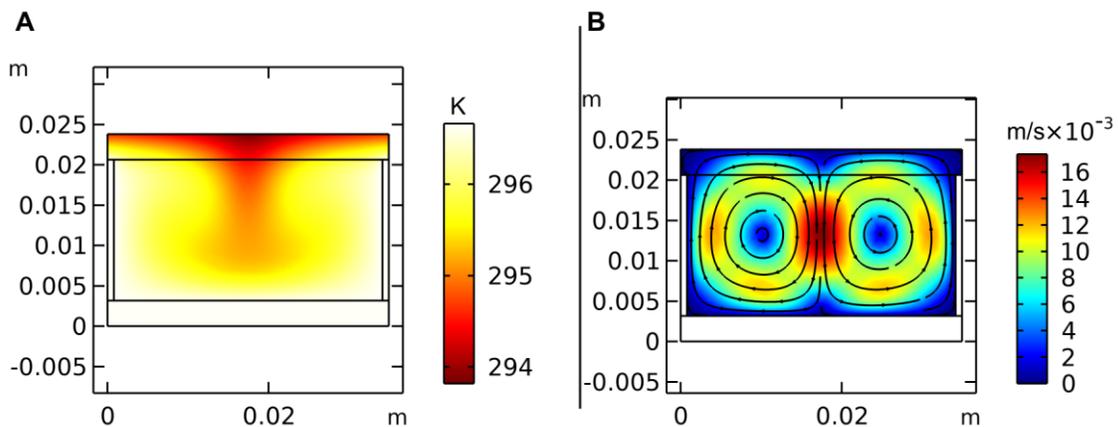

**Figure SI20: With Convection**
(A) Plot of the temperature distribution.
(B) Plot of the fluid velocity. Convection cells are clearly visible.

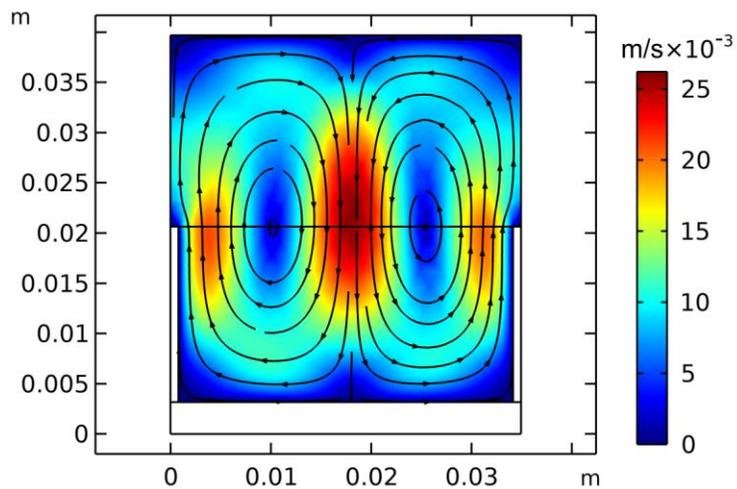

**Figure SI21: Convection cell in taller air gap**

Notably our simulations underestimates temperature. This is most evident in the flat geometry results in Table SI3, which predicts subambient radiative cooling. In our experiments, we have not observed sub-ambient radiative cooling during midday. We conducted our experiments by the coast and believe that using an atmospheric model that better accounts for the coastal conditions and humidity will mostly correct for the discrepancy.

Nonetheless, our simulations provide valuable insight. It shows that the heating of the louvers and box sidewalls are major heat sources and the convective shield results in convective cells forming which ultimately leads to the heating of the tiles. In any realistic implementation, the boundary effects will be minimal and the heat from the louvers will be convected away.

**Hysteresis and Switching Band Model Details**

*Overview*: To evaluate the impact of hysteresis and switching band size, we created a simple heat transfer model and evaluated the equilibrium temperature over the course of a day. We assume negligible thermal mass. The details are below.

For notational simplicity, all equations and constants have consistent units.

*Emissivity State Model*: A sigmoidal function, $S(T)$, describes how the state (i.e., emissivity) varies. We use a smooth function for this because the transitions are not instantaneous and discontinuous (i.e., louvers can open to an intermediate state). The sigmoid is centered about the switch temperature, $T_s$. The sigmoid function we use has the form:

$$S(T) = \frac{1}{1 + \exp\left(-k\left(T - T_s \pm \frac{\Delta T}{2}\right)\right)}$$

$k$ describes the sharpness of sigmoidal function. As $k \to \infty$, the sigmoidal function approaches a step function and as $k \to 0$, the function approaches a horizontal line.

For intuition, we define switching band, $b_s$, as the temperature difference between 10% and 90% switched. It is related to $k$ by the following equation.

$$b_s = -\frac{2\ln\left(\frac{1}{9}\right)}{k}$$

$\Delta T$ quantifies the hysteresis. It is the temperature offset of the heating and cooling curves. The cooling and heating sigmoid curves, $S_c$ and $S_h$, respectively, are:

$$S_c(T) = \frac{1}{1 + \exp\left(-k\left(T - T_s + \frac{\Delta T}{2}\right)\right)}$$

$$S_h(T) = \frac{1}{1 + \exp\left(-k\left(T - T_s - \frac{\Delta T}{2}\right)\right)}$$

$\epsilon_b$ and $\epsilon_w$ are the black and white state emissivities, respectively. They have both solar, $\epsilon_s$, and thermal, $\epsilon_t$, components. We assume the tile is spectrally flat within the solar and thermal radiation bands. For the black state, $\epsilon_s = 0.96$ and $\epsilon_t = 0.3$; for the white state, $\epsilon_s = 0.05$ and $\epsilon_t = 0.88$. These are based on the fabricated tile properties. The emissivity of the tile, $\epsilon$, along the heating sigmoid is described by the following relation:

$$\epsilon = \epsilon_b + \frac{\epsilon_w - \epsilon_b}{2} S_h(T)$$

Similar relationships can be written for the cooling curve and other state functions.

To go between the heating and cooling curves, the state remains constant until the heating or cooling sigmoid curve is reached. It then traces the heating or cooling state curve. See Fig. SI22 for further clarification.

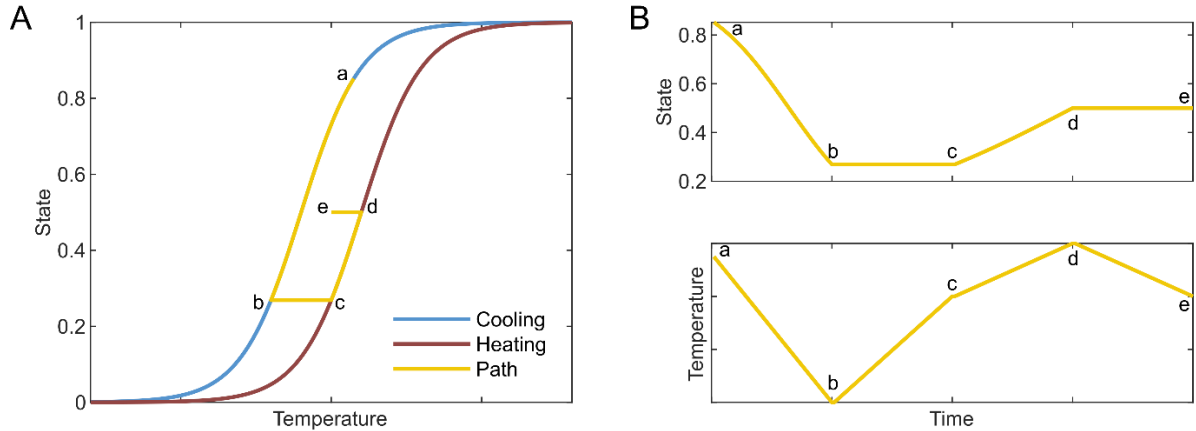

**Figure SI22: Heating and cooling path clarifying figure.**
(A) The path curve shows the state-temperature relationship for an adaptive surface undergoing thermal loading
(B) State and temperature vs time relationship for the path traced in A.

*Solar Flux Model*: For the solar flux model, we assume that there is 12 hours of sunlight a day and that it peaks at noon. The maximum solar flux is 1000 W/m². The equation that describes the solar heat gain, $P_s$, is

$$P_s = \begin{cases} -1000\epsilon_s \cos\left(\dfrac{2\pi t}{86400}\right), & \cos\left(\dfrac{2\pi t}{86400}\right) < 0 \\ \\ 0, & \cos\left(\dfrac{2\pi t}{86400}\right) \geq 0 \end{cases}$$

$t$ is the time in seconds. We initialize our simulations at midnight.

*Tile Radiative Model*: We assume that the tile emits radiation like a graybody. The radiative flux leaving the tile, $P_t$, is:

$$P_t = \sigma_b \epsilon_t T^4$$

$\sigma_b$ is the Stefan-Boltzmann constant.

*Ambient Temperature Model*: The ambient temperature, $T_a$, is assumed to vary sinusoidal between a cold temperature of $T_c$ and a high temperature of $T_h$. The temperature minimum occurs at 6:00 and the maximum temperature occurs at 18:00. The ambient temperature equation is:

$$T_a = \frac{T_c + T_h}{2} - \frac{T_h - T_c}{2}\sin\left(\frac{2\pi t}{86400}\right)$$

*Sky Radiation Model*: The radiative flux from the sky, $P_a$, is

$$P_a = \epsilon_t \int \epsilon_a(\lambda)\, I_{BB}(T_a, \lambda) d\lambda$$

$\epsilon_a$ is the atmospheric emittance. We use data from the Gemini observatory. For simplicity, we assume no angular dependence. $I_{BB}$ is the hemispherical blackbody intensity. It is a function of the ambient temperature and wavelength, $\lambda$.

*Convection Model*: We assume a constant convection coefficient. For natural convection, a convection coefficient of about $h = 25 \frac{W}{m^2{}^\circ C}$ is typical[2]. The convective power, $P_c$, is

$$P_c = h(T - T_a)$$

*Implementation*: To determine the equilibrium temperature, we solve the following root finding problem numerically:

$$P_s + P_a - P_c - P_t = 0$$

For each time step, we record the state value, $s$, and the equilibrium temperature, $T$.

To quantify the performance of different cases, we compute the variance, $v$, from the switch temperature.

$$v = \sum \frac{(T - T_s)^2}{n}$$

$n$ is the number of steps evaluated. For cases with hysteresis, we remove the first 24 hours to eliminate the transience.